\documentclass[11pt,a4paper]{article}

\usepackage{amsfonts}
\usepackage{multirow}
\usepackage{graphicx}
\usepackage{float}

\title{\textbf{Spectral Flow of Vortex Shape Modes over \\
    the BPS 2-Vortex Moduli Space}}

\author{A. Alonso Izquierdo$^{(a)}$, W. Garcia Fuertes$^{(b)}$,
N. S. Manton$^{(c)}$ and J. Mateos Guilarte$^{(d)}$
\\ {\normalsize {\it $^{(a)}$ Departamento de Matematica
Aplicada}, {\it Universidad de Salamanca, SPAIN}}\\ 
{\normalsize {\it $^{(b)}$ Departamento de Fisica}, {\it Universidad de Oviedo, SPAIN}} \\ 
{\normalsize {\it $^{(c)}$ Department of Applied Mathematics and Theoretical Physics}, {\it University of Cambridge, UK}} \\ 
{\normalsize {\it $^{(d)}$ Departamento de Fisica Fundamental}, {\it
    Universidad de Salamanca, SPAIN}}}

\date{}

\begin{document}

\maketitle

\begin{abstract}
The flow of shape eigenmodes of the small fluctuation operator around BPS
2-vortex solutions is calculated, as a function of the intervortex separation
$2d$. For the rotationally-invariant 2-vortex, with
$d = 0$, there are three discrete modes; the lowest is
non-degenerate and the upper two are degenerate. As $d$ increases,
the degeneracy splits, with one eigenvalue increasing
and entering the continuous spectrum, and the other decreasing and
asymptotically coalescing with the lowest eigenvalue, where they jointly
become the eigenvalue of the 1-vortex radial shape mode. The behaviour
of the eigenvalues near $d=0$ is clarified using a perturbative
analysis, and also in light of the 2-vortex moduli space geometry.

\end{abstract}


\section{Introduction}

Collective coordinate dynamics has been successfully implemented in
\cite{MaOlRoWe} to describe the low-energy effective theory governing
collisions between $\phi^4$-kinks. The collective coordinates associated
to each kink consist of the kink center $a$ and the amplitude $A$ of
the unique shape mode of kink fluctuations. The reduced system in the
centre of mass frame, assuming reflection symmetry, has a two-dimensional
Lagrangian with a kinetic energy involving a metric depending on $A$ and
a potential energy also depending on $A$.

It seems feasible to develop a similar effective theory for BPS $n$-vortex
dynamics in the Abelian Higgs model at critical coupling. For a 1-vortex, the
collective coordinates would be the vortex center and the amplitude of
the unique discrete fluctuation mode -- the radial shape mode. Finding a
collective coordinate treatment in the case of 2-vortices would
be far more interesting. Here, the discrete eigenvalues and eigenfunctions
of the second-order fluctuation operator vary with the
separation of the 1-vortex constituents. Internal shape modes
of vortices of various kinds have been studied, e.g., in refs.
\cite{AlGarGuil, AlGarGuil2, Hindmarsh, Arodz1}, but so far, almost
exclusively for coincident-vortex solutions.

The moduli space of BPS $n$-vortices is the space of sets
of $n$ unordered points in the plane -- the points where the Higgs field
vanishes. This is a non-singular complex manifold \cite{Tau,Samols,ManSut},
even where vortices coincide, but one needs to be careful
about the choice of coordinates. In particular, for 2-vortices with fixed
centre of mass at the origin, if the constituent vortices have a
separation $2d$, then it is $d^2$ rather than $d$ that is a good
radial coordinate on the moduli space.

For each $n>0$ there is a unique $n$-vortex, rotationally-invariant
around the origin, for which the $n$ constituent vortex locations coincide.
The spectral problem of the second-order vortex fluctuation operator
in this case resembles that of the planar Hydrogen atom, but the potential
well is bounded from below and reaches the continuous-spectrum threshold
exponentially fast as the radius increases, which implies that there exist
at most a finite number of discrete shape modes. In \cite{AlGarGuil}, these
facts guided the search for discrete modes. It was found, for example, that
for $n=1$ there is only one discrete shape mode whereas for $n=2$ there
are three.

The generic BPS 2-vortex is formed from two $1$-vortices separated
by an arbitrary finite distance. The spectrum of vortex fluctuations is
then akin to the quantum spectrum of a particle moving in the field
created by two centers of force, where the centers are freely movable.
In this paper we investigate how the shape modes and their eigenfrequencies
vary with the vortex separation. The number of discrete modes remains
finite, and in fact decreases from three to two as the separation increases.

Our results should lead to a
generalisation of the notion of geodesic flow on the 2-vortex moduli
space as an approximation to 2-vortex dynamics. Allowing for the
possibility of excited shape modes will require a more complicated
collective coordinate model for the low-energy dynamics of vortices,
including a potential on the moduli space. If the vortex motion is
slow, and the evolution of the shape mode amplitudes is treated
adiabatically, then a Berry connection will probably also be needed.

\section{BPS vortices in the Abelian Higgs model:
SUSY structure of the fluctuation operator}

We start from the action of the relativistic Abelian Higgs model, describing
the minimal coupling between a $U(1)$ gauge field and a charged scalar
field in a phase where the gauge symmetry is broken spontaneously. We
focus on the BPS critical value of the coupling (the
strength of the Higgs potential) where the Higgs and gauge field
masses are equal \cite{Bogomolny, Prasad}. In terms of non-dimensional
coordinates, couplings and fields, the action functional for this
system is
\begin{equation}
S[\phi,A]=\int dx_0dx_1dx_2 \Big[ -\frac{1}{4} F_{\mu\nu}F^{\mu \nu} +
\frac{1}{2} \overline{D_\mu \phi}\, D^\mu \phi -\frac{1}{8} (\overline{\phi}\, \phi-1)^2 \Big]
\, .
\label{action1}
\end{equation}
The ingredients here are a complex scalar (Higgs) field
$\phi(x)=\phi_1(x)+i\phi_2(x)$, a $U(1)$ gauge potential
$A_\mu(x)=(A_0(x),A_1(x),A_2(x))$, the covariant derivative $D_\mu
\phi(x) = (\partial_\mu -i A_\mu(x))\phi(x)$ and the electromagnetic
field tensor $F_{\mu\nu}(x)=\partial_\mu A_\nu(x) - \partial_\nu
A_\mu(x)$. Our Minkowski-space metric tensor is
$g_{\mu\nu}={\rm diag}(1,-1,-1),$ with $\mu,\nu=0,1,2$, and we use
the Einstein summation convention. In the temporal gauge $A_0=0$,
the energy of static field configurations becomes
\begin{equation}
V[\phi,A]= \int_{\mathbb{R}^2} d^2x \Big[ \frac{1}{2} F_{12}^2 +
\frac{1}{2} \overline{D_1\phi}\, D_1\phi + \frac{1}{2} \overline{D_2\phi}\,
D_2\phi+\frac{1}{8} (\overline{\phi}\,\phi-1)^2 \Big] \, .
\label{energyfunctional}
\end{equation}
We interchangably use Cartesian and polar coordinates
$\vec{x} = (x_1,x_2) = (x,y) = (r\cos\theta, r\sin\theta)$ with $d^2x
= dx \, dy$.
The energy (\ref{energyfunctional}), treated non-relativistically, models the
free energy of a superconducting material arising in the Ginzburg--Landau
theory of superconductivity -- see formula (17) in \cite{Abrikosov} where the
order parameter $|\phi|^2 = \overline{\phi} \, \phi$ corresponds to
the density of Cooper pairs.

Critical points of $V[\phi,A]$ that satisfy the boundary conditions 
\begin{equation}
\overline{\phi}\, \phi|_{S_\infty^1}=1 \, , \quad 
D_i\phi|_{S_\infty^1}=0 \quad \mbox{and} \quad
F_{12}|_{S_\infty^1}=0 \, \,
\label{asymptotic}
\end{equation}
on the circle at infinity $S_\infty^1$ (i.e. as $r \rightarrow \infty$)
have finite energy. Indeed, it can be checked that the configuration
space of static fields,
\begin{equation}
{\cal C}=\left\{ \{\phi, A \} \ {\rm s.t.} \ V[\phi,A] < \infty \right\}
=\sqcup_{n\in\mathbb{Z}}\,\,  {\cal C}_n \, ,
\end{equation}
is the union of $\mathbb{Z}$ topologically disconnected sectors. Here,
$n$ is the vorticity or winding number of the map
$\phi \vert_\infty: S^1_\infty \, \, \longrightarrow  \, \, S^1$
from the circle at infinity to the vacuum orbit $|\phi|^2 = 1$,
parametrised by the phase of $\phi$. It follows from
the vanishing of the covariant derivative of $\phi$ at infinity, that
$n$ is also the normalised magnetic flux, $\Phi \equiv \frac{1}{2\pi}
\int_{\mathbb{R}^2} d^2 x \, F_{12} = n$. In the BPS regime, $V[\phi,A]$
can be written in the form, see \cite{Bogomolny},
\begin{equation}
V[\phi,A]=\frac{1}{2}\int_{\mathbb{R}^2} d^2 x
\Big[\Big(F_{12}\pm\frac{1}{2}(\overline{\phi}\,\phi-1)\Big)^2+\left\vert
  D_1\phi\pm iD_2\phi\right\vert^2\Big] \pm
\frac{1}{2}\int_{\mathbb{R}^2} d^2 x \,F_{12} \, ,
\end{equation}
which implies that BPS vortices are solutions of the first-order PDEs 
\begin{equation}
D_1\phi \pm i D_2 \phi =0 \, , \hspace{0.5cm} F_{12}\pm
\frac{1}{2} (\overline{\phi}\,\phi -1)=0 \, ,
\label{fopdes}
\end{equation}
and are absolute minima of the energy for each $n$, with $V[\phi,A]=\pi
\vert n\vert$. For $n$ positive, the upper signs in (\ref{fopdes})
need to be chosen. Then, given $n$, there exist BPS vortex solutions
characterized by $n$ arbitrary locations (points) in the plane, which are 
the zeros of the scalar field counted with multiplicity, and
simultaneously the locations of maximal magnetic field. $n$-vortex
solutions therefore have $2n$ real moduli. For the sake of clarity
in later formulas we shall denote the scalar field profile
$\phi(\vec{x})$ of an $n$-vortex solution as $\psi^{(n)}(\vec{x})
= \psi_1^{(n)}(\vec{x}) + i \, \psi_2^{(n)}(\vec{x})$ and the gauge
(vector) potential $A_k(\vec{x})$ as $V_k^{(n)}(\vec{x})$. The main theme
of this paper is the construction of field fluctuations around
2-vortices $\{\psi^{(2)}(\vec{x}), V_k^{(2)}(\vec{x}) \}$, and
analysis of how they depend on the vortex separation.

To find the linear fluctuation modes, we consider the evolution
of small perturbations $\varphi_j(\vec{x})$ and $a_k(\vec{x})$ around
BPS vortex fields $\psi_j^{(n)}(\vec{x})$ and $V_k^{(n)}(\vec{x})$.
The total fields are
\begin{equation}
\phi_j(\vec{x}) = \psi_j^{(n)}(\vec{x}) + \epsilon\,
\varphi_j(\vec{x}) \, e^{i \omega t} \, , \hspace{0.5cm}
A_k(\vec{x}) = V_k^{(n)}(\vec{x}) + \epsilon \, a_k(\vec{x})\, e^{i\omega t}
\label{perturbed}
\end{equation}
with $\epsilon$ small, and they still belong to the $n$-vortex
topological sector. To discard pure gauge fluctuations, we impose the
\textit{background gauge}
\begin{equation}
B(\varphi_j,a_k)=\partial_k a_k( \vec{x})-(\,\psi_1^{(n)}( \vec{x})\,
\varphi_2( \vec{x})-\psi_2^{(n)}( \vec{x})\,\varphi_1( \vec{x})\,)=0
\label{backgroundgauge}
\end{equation}
as the gauge fixing condition on the perturbations. Substituting
(\ref{perturbed}) into the field equations and linearizing, the
eigenfrequencies $\omega$ and eigenmodes
\begin{equation}
\xi(\vec{x})=\left( \begin{array}{c c c c}a_1(\vec{x}) \,, & a_2(\vec{x})
\,, & \varphi_1(\vec{x}) \,, & \varphi_2(\vec{x}) \end{array} \right)^t
\end{equation}
are found to be solutions of the spectral problem ${\cal H}^+
\xi_\lambda(\vec{x}) =\omega_\lambda^2 \, \xi_\lambda(\vec{x})$, see
\cite{guilarte0, guilarte, guilarte1}. Here, $\lambda$ is a label in
either the discrete or continuous spectrum, and ${\cal H}^+$
is the second-order vortex fluctuation operator
\begin{equation}
{\cal H}^+= \left( \begin{array}{cccc}
-  \nabla^2 + |\psi|^2 & 0 & -2D_1 \psi_2 & 2 D_1 \psi_1 \\
0 & -  \nabla^2 +|\psi|^2 & -2 D_2 \psi_2 & 2 D_2 \psi_1 \\
-2 D_1 \psi_2 & -2 D_2\psi_2 & - \nabla^2 +\frac{1}{2}(3|\psi|^2-1)+V_kV_k
& -2 V_k \partial_k -\partial_k V_k \\
2D_1\psi_1 & 2 D_2 \psi_1 & 2V_k \partial_k + \partial_k V_k
& - \nabla^2 +\frac{1}{2} (3|\psi|^2-1) + V_kV_k
\end{array} \right)
\label{soflucdo}
\end{equation}
where $|\psi|^2 = \psi_1^2 + \psi_2^2$, and $\{\psi, V_k \}$ satisfy
eqs.(\ref{fopdes}). The fluctuation vectors $\xi(\vec{x})$ belong in
general to a rigged Hilbert space. There are square integrable eigenfunctions
$\xi_\lambda(\vec{x})\in L^2(\mathbb{R}^2)\otimes \mathbb{R}^4$ belonging to
the discrete spectrum, for which the squared norm $\|\xi_\lambda(\vec{x})\|^2
= \int_{\mathbb{R}^2} d^2x [ (a_1(\vec{x}))^2 + (a_2(\vec{x}))^2
+ (\varphi_1(\vec{x}))^2 + (\varphi_2(\vec{x}))^2 ] $ is
bounded, as well as continuous spectrum eigenfunctions.

Weinberg \cite{Weinberg} proved that there are $2n$ linearly-independent
normalizable zero modes (having eigenvalues $\omega^2 = 0$) for any BPS
$n$-vortex solution. These can be characterized as lying in the kernel
of the operator
\begin{equation}
{\cal D}= \left( \begin{array}{cccc}
-\partial_2 & \partial_1 & \psi_1 & \psi_2 \\
-\partial_1 & -\partial_2 & -\psi_2 & \psi_1 \\
\psi_1 & -\psi_2 & -\partial_2 + V_1 & -\partial_1 -V_2 \\
\psi_2 & \psi_1 & \partial_1+V_2 & -\partial_2 + V_1
\end{array} \right)  \label{zeromoded} \, \, .
\end{equation}
Analysis of these zero modes was further developed in \cite{Ruback}
and \cite{Burzlaff}, motivated by the study of vortex
scattering at low energies within the approach of geodesic dynamics in
the $n$-vortex moduli space, see e.g. \cite{guilarte2}.

A crucial point for the calculations in this paper is that
\begin{equation}
{\cal H}^+={\cal D}^\dagger \, {\cal D} \qquad \mbox{and}
\qquad {\cal H}^-={\cal D} \, {\cal D}^\dagger \label{susy01}
\end{equation}
are SUSY (supersymmetry) partners, and therefore isospectral in the strictly
positive part of the spectrum. Moreover,
\begin{equation}
{\cal H}^- =  \left( \begin{array}{cccc}
- \nabla^2 + |\psi|^2 & 0 & 0 & 0 \\
0 & -  \nabla^2 +|\psi|^2 & 0 & 0 \\
0 & 0 & - \nabla^2 +\frac{1}{2} (|\psi|^2+1)+V_kV_k &
-2 V_k \partial_k -\partial_k V_k \\
0 & 0 & 2V_k \partial_k + \partial_k V_k & - \nabla^2 +\frac{1}{2}
(|\psi|^2+1) + V_kV_k \end{array} \right)
\end{equation}
is a simpler operator than ${\cal H}^+$, and its spectrum is easier to
investigate. For ${\cal H}^-$, it was proved in \cite{AlGarGuil,AlGarGuil2}
that it is sufficient to find eigenfunctions of the form
$\xi_\lambda^{\rm -}(\vec{x})=(a_1(\vec{x}) \, , 0 \, , 0 \, , 0 )^t$,
satisfying
\begin{equation}
(-  \nabla^2 + |\psi^{(n)}(\vec{x})|^2)\, a_1(\vec{x})= \omega_\lambda^2
\, a_1(\vec{x})
\label{pdeA} \, .
\end{equation}
The corresponding eigenfunction of the SUSY partner operator
${\cal H}^+$, sharing the positive eigenvalue $\omega_\lambda^2$ and
the same normalization, is
\begin{equation}
\xi^+_\lambda(\vec{x})= \frac{1}{\omega_\lambda } {\cal D}^\dagger
\xi_\lambda^-(\vec{x}) = \frac{1}{\omega_\lambda }
\left( \begin{array}{cccc} \partial_2 a_1(\vec{x}) \, ,
& -\partial_1a_1(\vec{x}) \, , & \psi_1(\vec{x})a_1(\vec{x}) \, ,
& \psi_2(\vec{x}) a_1(\vec{x}) \end{array}\right)^t .
\label{autofuncion1}
\end{equation}

\section{The spectrum of the BPS 2-vortex fluctuation
operator}

In this Section we obtain numerically the discrete, positive eigenvalues
and eigenfunctions of the second-order fluctuation operator ${\cal H}^+$
evaluated at BPS 2-vortex solutions $\{\psi^{(2)}(\vec{x}),
V_k^{(2)}(\vec{x})\}$. Recall that these 2-vortices can be interpreted as two
separated $1$-vortices. If the 2-vortex mass center is located at the
origin, the $1$-vortex locations (zeros of $\psi^{(2)}$) can be
assumed to be at $(d,0)$ and $(-d,0)$, with equivalent solutions
being obtained by translation and rotation. The first task is to construct
the 2-vortex solution with these zeros.

We start from the more general, rotationally-invariant $n$-vortex
solution, having the polar coordinate form
\begin{equation}
\psi^{(n)}(\vec{x}) = f_n(r) \, e^{in\theta} \, ,
\hspace{0.5cm} r A^{(n)}_\theta(\vec{x}) = n\, \beta_n (r) \, ,
\end{equation}
and with the radial gauge $A_r^{(n)}=0$ imposed. The
functions $f_n(r)$ and $\beta_n(r)$ need to satisfy the first-order
coupled equations
\begin{equation}
\frac{df_n}{dr} = \frac{n}{r} f_n(r) [1-\beta_n(r)]
\, , \hspace{0.5cm} \frac{d \beta_n}{dr} = \frac{r}{2n} [1-f_n^2(r)]
\, ,
\label{edo1}
\end{equation}
and the asymptotic conditions $f_n(r)\rightarrow 1$ and $\beta_n(r)
\rightarrow 1$ as $r\rightarrow \infty$. The requirement of
regularity at the origin fixes the behaviour for small
$r$ to be $f_n(r) \sim d_n r^n$ and $\beta_n(r) \sim \frac{1}{4n} r^2$, where
$d_n$ is a constant depending on $n$. Using eqs.(\ref{edo1}),
the radial profiles $f_n(r)$ and $\beta_n(r)$ for $n=1$ and $n=2$
(and higher $n$) are easily generated. A $2$-vortex whose zeros are well
separated is then approximated by superposing two rotationally-invariant
1-vortices, translated to have the desired zeros. A 2-vortex with
coincident zeros is the rotationally-invariant solution with radial
profiles $f_{2}(r)$ and $\beta_{2}(r)$.

To construct $2$-vortex solutions whose zeros are separated by an intermediate
distance $2d$, we take advantage of the Bogomolny energy bound, which for
2-vortices is
$V[\psi^{(2)}, V^{(2)}] = 2 \pi$. Our strategy is to numerically construct
an $n=2$  configuration with zeros at $(d,0)$ and $(-d,0)$, that
accurately saturates this bound. Indeed, the deviation of the energy
from this bound is an estimate of the solution's precision. To generate
initial data, we use a generalized $n=1$ configuration centered at $(d,0)$,
\begin{eqnarray}
&& \phi^{(1)}(\vec{x},d,\alpha) = \Big[ \alpha f_{1}(\overline{r})  +
   (1-\alpha) \sqrt{f_{2} (\overline{r})} \Big]
   e^{i\overline{\theta}} \, , \nonumber \label{confn1a} \\ 
&& A_1^{(1)} (\vec{x},d,\alpha)= - \Big[ \alpha \beta_{1}(\overline{r})
   + (1-\alpha) \beta_{2} (\overline{r}) \Big]
 \frac{\sin \overline{\theta}}{\overline{r}} \, , \nonumber \label{confn1b} \\
&& A_2^{(1)} (\vec{x},d,\alpha) =   \Big[ \alpha \beta_{1}(\overline{r})
   + (1-\alpha) \beta_{2} (\overline{r}) \Big]
   \frac{\cos \overline{\theta}}{\overline{r}} \, , \label{confn1c} 
\end{eqnarray}
where $\overline{r}=\sqrt{(x- d)^2 + y^2}$ and $\overline{\theta}=
\arctan \frac{y}{x-d}$ are polar coordinates around the centre.
Here, $\alpha$ is a free parameter whose value is chosen later. The
desired $n=2$ configuration can now be constructed using the standard
superposition
\begin{eqnarray}
&& \phi^{(2)}(\vec{x},d,\alpha) = \phi^{(1)}(\vec{x},d,\alpha) \cdot
   \phi^{(1)}(\vec{x},-d,\alpha) \, , \nonumber \\
&& A_k^{(2)} (\vec{x},d,\alpha) = A_k^{(1)} (\vec{x},d,\alpha)
   + A_k^{(1)} (\vec{x},-d,\alpha) \, .
\label{productansatz}
\end{eqnarray} 
Substituting this into (\ref{energyfunctional}) gives the energy
\begin{equation}
V(\alpha) = V[\phi^{(2)}(\vec{x},d,\alpha),A^{(2)}(\vec{x},d,\alpha)]
\, ,
\label{funalpha}
\end{equation}
depending on $\alpha$. Our numerical mesh ranges over the spatial rectangle
$[-30,30]\times [-15,15]$, with $I_{\rm max}=3201$ points in the
$x$-component and $J_{\rm max}=1601$ in the $y$-component.
The mesh points are $\vec{p}_{ij} = (x_{\rm min} + i\,\delta x
, y_{\rm min} + j \,\delta y  )$ where $x_{\rm min}=-30$ and $y_{\rm min}=-15$,
$\delta x= \delta y = 0.01875$ are the spatial steps, $i=0,\dots, I_{\rm
  max}-1$ and $j=0,\dots, J_{\rm max}-1$. A second-order finite
difference scheme is employed to approximate the spatial derivatives
arising in the functional energy (\ref{funalpha}). Now, for each $d$,
the value of $\alpha$ is chosen to minimize (\ref{funalpha}). The analytical
relation $\alpha = d^\epsilon/(d^\epsilon+ \gamma)$ with $\epsilon=3.4$ and
$\gamma=1.4$ is a good approximation. Finally, from this configuration,
numerical gradient flow is employed to refine the solution. The resulting
energies saturate the Bogomolny bound with a relative error less
than 0.03\%. These numerical 2-vortex solutions are precise enough
to address the spectral problem of this paper. For later convenience,
we denote their scalar field at the mesh points by
$\psi^{(2)}(\vec{p}_{ij})$.

The next task involves the spatial discretization of the spectral problem
(\ref{pdeA}) for $n=2$, which will lead us to the positive
discrete spectrum of the 2-vortex fluctuation operator ${\cal H}^-$,
and hence ${\cal H}^+$. Here, we can work with a less fine mesh than the
previous one. We introduce new mesh points
$\vec{q}_{ij} = (x_{\rm min} + i \Delta x , y_{\rm min} + j\Delta y)$
where $\Delta x$ and $\Delta y$ are the new spatial steps in each
direction,  $i=0,\dots,i_{\rm max}-1$ and $j=0,\dots,j_{\rm
  max}-1$. The fluctuation field values at the mesh points,
$(a_1)_{ij} = a_1(\vec{q}_{ij})$, are arranged in a single
column $(a_1)_s \equiv ((a_1)_{00} \, (a_1)_{01} \, \dots \,
(a_1)_{0(j_{\rm max}-1)} \, (a_1)_{10} \, \dots \,
(a_1)_{(i_{\rm max}-1)(j_{\rm max}-1)})^t$ where
$s= j_{\rm max} \cdot i + j \, , $
as also are the values of the background potential well,
\begin{equation}
U(\vec{q}_s) \equiv U (\vec{q}_{ij}) =
| \psi^{(2)}(\vec{q}_{ij}) |^2 \, .
\end{equation}
With this arrangement, the
eigenvalue problem, discretized up to second order, is
\begin{equation}
- \frac{(a_1)_{s+j_{\rm max}} - 2 (a_1)_{s} + (a_1)_{s-j_{\rm max}}}
{(\Delta x)^2} - \frac{(a_1)_{s+1} - 2 (a_1)_{s} + (a_1)_{s-1}}{(\Delta y)^2}
+ U (\vec{q}_s)  \, (a_1)_{s} = \omega_\lambda^2 (a_1)_{s} \, .
\label{discrspec}
\end{equation}

Dirichlet boundary conditions have been assumed for the eigenfunctions,
which implies that the value $(a_1)_{s}=0$ is imposed for $s<0$
and for $s\geq j_{\rm max} \cdot i_{\rm max}$. In addition, if
$s \,{\rm mod} \, j_{\rm max} =0$ then
$(a_1)_{[s/j_{\rm max}] j_{\rm max} \, - \, 1}=0$ where $[z]$ denotes the
integer part of $z$, and likewise, $(a_1)_{[s/j_{\rm max}] j_{\rm max}}=0$
for $s \,{\rm mod} \, j_{\rm max} = j_{\rm max}-1$.
The procedure approximates the fluctuation operator by a
finite matrix, which can be analysed to obtain the full spectrum of
discrete eigenvalues and eigenfunctions. It has been checked that the choice
$i_{\rm max}=201$ and $j_{\rm max}=101$ with $\Delta x = \Delta y = 0.3$,
giving a fluctuation operator approximated by a $20301 \times 20301$
matrix, provides precise enough results. In Figure 1, the potential
wells $U(\vec{q}_s)$ of the discrete Schr\"odinger-type equation
(\ref{discrspec}) are plotted for intervortex distance
parameters $d=0$, $d=2$ and $d=5$.

\begin{figure}[h]
	\centering
	\includegraphics[height=3cm]{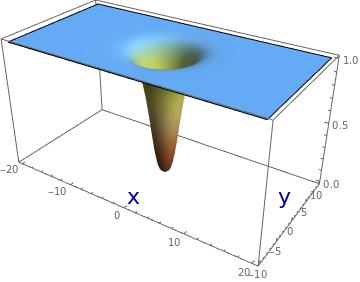} \hspace{1cm}
	\includegraphics[height=3cm]{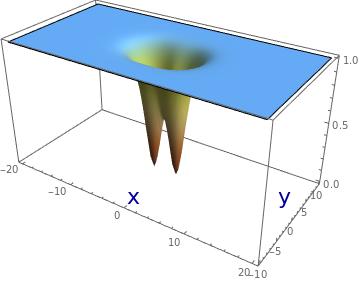} \hspace{1cm}
	\includegraphics[height=3cm]{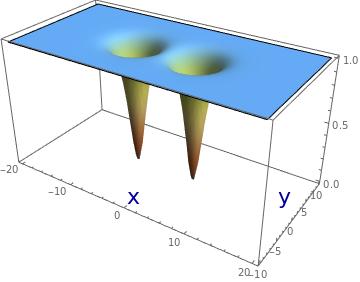}
	\caption{Potential wells $U = |\psi^{(2)}|^2$ of the 2-vortex
 fluctuation operator for $d=0$ (left), $d=2$ (middle) and $d=5$ (right).}  
	\label{fig:potentialwells}
\end{figure}

\begin{figure}[h]
	\centering
	\includegraphics[height=4.5cm]{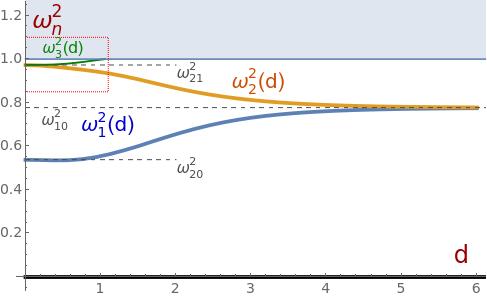} \hspace{0.5cm}
	\includegraphics[height=4.5cm]{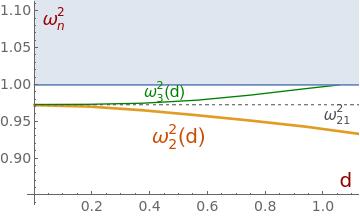}
	\caption{Eigenvalues of the 2-vortex fluctuation
          operator plotted against the intervortex separation parameter $d$
          (left). The red boxed area in the left figure has been
          enlarged on the right.}
        \label{fig:spectrum2}
\end{figure}

The main result of this paper is displayed in Figure \ref{fig:spectrum2}.
This shows the three discrete positive eigenvalues $\omega_\lambda^2$, plotted
as functions of $d$. To better understand the result we recall the spectrum
of the fluctuation operator in the case of rotationally-invariant 1- and
2-vortices. Using the notation introduced in \cite{AlGarGuil, AlGarGuil2},
for the 1-vortex there is only one shape eigenmode with angular momentum
number $k=0$, eigenvalue $\omega_{10}^2 \approx 0.777476$ and
eigenfunction $v_{10}(r)$. For the 2-vortex there is one eigenmode with
angular momentum number $k=0$, eigenvalue $\omega_{20}^2 \approx
0.53859$ and eigenfunction $v_{20}(r)$, and also a doubly-degenerate
pair of eigenmodes with angular momentum number $k=1$, eigenvalue
$\omega_{21}^2 \approx 0.97303$ and eigenfunctions
$v_{21}(r)\cos \theta$ and $v_{21}(r)\sin\theta$. The behaviour of the
radial profiles $v_{10}(r)$, $v_{20}(r)$ and $v_{21}(r)$ is shown in
refs.\cite{AlGarGuil, AlGarGuil2}. We emphasize that the $k=1$ mode is
doubly degenerate when $d=0$, because of the rotational invariance of
the 2-vortex.

As the two vortices separate, the low-lying $k=0$ eigenvalue
$\omega_1^2(d)$ monotonically increases from the value $\omega_1^2(0)
= \omega_{20}^2$ to $\omega_1^2(\infty) = \omega_{10}^2$, and the degeneracy
of the next two eigenmodes is broken. The $k=1$ eigenvalue splits into
$\omega_{2}^2(d)$ and $\omega_{3}^2(d)$. The upper eigenvalue
$\omega_{3}^2(d)$ increases from $\omega_{3}^2(0) = \omega_{21}^2$
as $d$ increases, rapidly reaching
the threshold of the continuous spectrum close to $d=1$ where it
disappears. On the other hand, $\omega_{2}^2(d)$ decreases
from $\omega_{2}^2(0)= \omega_{21}^2$ and fuses with
$\omega_{1}^2(d)$ around $d=5$, where both approach
$\omega_{10}^2$. More precisely, for two asymptotically-separated 1-vortices
there are doubly-degenerate eigenmodes with eigenvalue $\omega_{1}^2(\infty)=
\omega_{2}^2(\infty) = \omega_{10}^2$. These modes correspond to the
symmetric and antisymmetric combinations of the localized, radial
shape modes associated to the individual 1-vortices.

\begin{figure}[h]
	\centering
	\includegraphics[height=4cm]{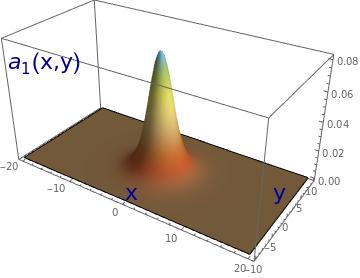} \hspace{0.5cm}
	\includegraphics[height=4cm]{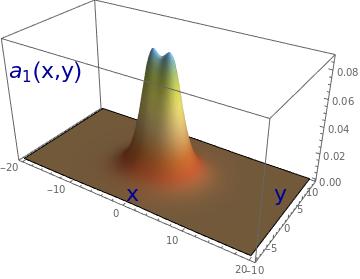} \hspace{0.5cm}
	\includegraphics[height=4cm]{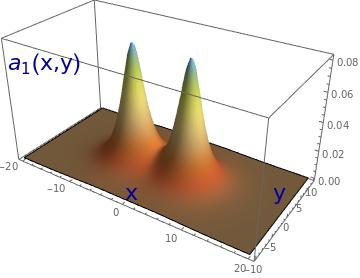}
	\caption{Eigenmode of the 2-vortex fluctuation operator
          with eigenvalue $\omega_1^2(d)$ for $d=0$ (left), $d=2$ (middle) and
          $d=5$ (right).}
        \label{fig:eigenfunctions1}
\end{figure}

\begin{figure}[h]
	\centering
	\includegraphics[height=4cm]{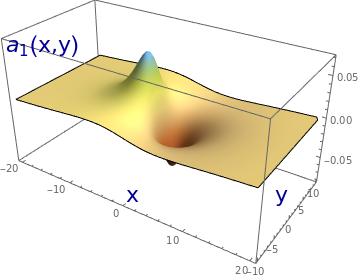} \hspace{0.5cm}
	\includegraphics[height=4cm]{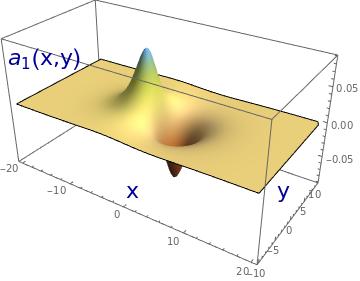} \hspace{0.5cm}
	\includegraphics[height=4cm]{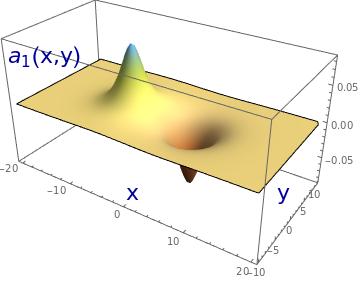}
	\caption{Eigenmode of the 2-vortex fluctuation operator with
          eigenvalue $\omega_2^2(d)$ for $d=0$ (left), $d=2$ (middle) and
          $d=5$ (right).}
        \label{fig:eigenfunctions2}
\end{figure}

To emphasize this picture, in Figure \ref{fig:eigenfunctions1} the
eigenmode $a_1$ of ${\cal H}^-$ with eigenvalue $\omega_1^2(d)$ is plotted for
selected values of $d$. We observe how starting at $d=0$
from the $k=0$ solution $a_1(r)=v_{20}(r)$, the eigenmode
remains symmetric. Asymptotically, it may be understood as the
symmetric linear combination of the 1-vortex modes, $a_1\simeq
v_{10}(\overline{r}_+)+v_{10}(\overline{r}_-)$ with $\overline{r}_\pm
=\sqrt{(x \pm d)^2 + y^2}$. In Figure \ref{fig:eigenfunctions2} we
plot the eigenmode with eigenvalue $\omega_2^2(d)$, starting from
the $k=1$ mode $a_1=v_{21}(r)\cos\theta$ that is antisymmetric under
$x \to -x$. The mode profile retains its antisymmetry as $d$ grows. In
particular, for large $d$, the mode
becomes the antisymmetric linear combination of the 1-vortex
modes, $a_1\simeq v_{10}(\overline{r}_+)-v_{10}(\overline{r}_-)$. 

\begin{figure}[h]
	\centering
		\includegraphics[height=2.5cm]{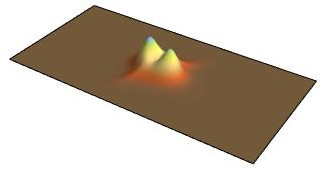} 
		\includegraphics[height=2.5cm]{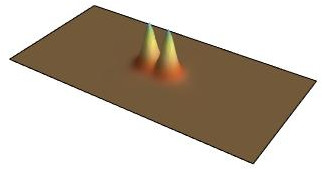} 
		\includegraphics[height=2.5cm]{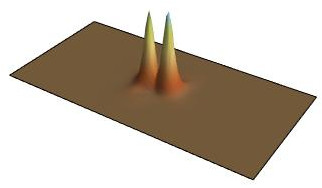} 
	
	\caption{Snapshots of the oscillating 2-vortex at
          $d=1.5$, excited by the mode of frequency $\omega_1$.}
        \label{fig:energies1}
\end{figure}

\begin{figure}[h]
	\centering
		\includegraphics[height=2.5cm]{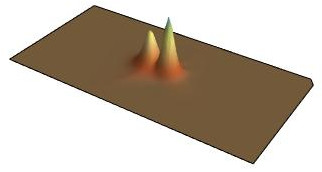} 
		\includegraphics[height=2.5cm]{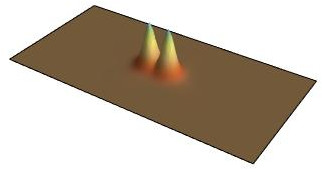} 
		\includegraphics[height=2.5cm]{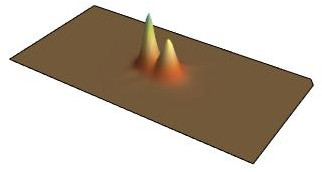}
	
	\caption{Snapshots of the oscillating 2-vortex at $d=1.5$
          excited by the mode of frequency $\omega_2$.}
        \label{fig:energies2}
\end{figure}

The eigenmodes of the 2-vortex fluctuation operator ${\cal H}^+$ can
be obtained from these eigenmodes of ${\cal H}^-$ by using
the intertwining formula (\ref{autofuncion1}). Since these are
4-component vectors, it is difficult to illustrate their
precise form, or the way that they excite a 2-vortex solution.
However, we can plot the potential energy density of an excited
2-vortex. Figure \ref{fig:energies1} shows snapshots
of the oscillating 2-vortex solution at separation parameter $d=1.5$,
excited by the mode of lowest positive frequency, $\omega_1$. The
constituent 1-vortices shrink and stretch in phase. Figure
\ref{fig:energies2} shows the oscillations of the same 2-vortex
excited by the discrete mode with the higher frequency, $\omega_2$.
In this case, the 1-vortices oscillate in counterphase, i.e. while
one vortex shrinks the other stretches.

\section{Perturbation theory near $d=0$}
\label{Pert}

Here, we present some analytical calculations based on
perturbation theory, that explain the spectral structure of the
2-vortex modes near the rotationally-invariant 2-vortex at $d=0$.
Specifically, we shall consider the eigenvalue problem (\ref{pdeA}) where the
potential well is determined by the scalar part of a 2-vortex solution
for small $d$, which we denote by
\begin{equation}
\widetilde{\psi}^{(2)}(\vec{x}) = \psi^{(2)}(\vec{x}) + \epsilon \,
\delta\psi^{(2)}(\vec{x}) + \cdots \, .
\label{expansion1}
\end{equation}
$\psi^{(2)}(\vec{x})$ is the rotationally-invariant 2-vortex
scalar field and $\delta\psi^{(2)}(\vec{x})$ is derived from the
zero-frequency mode $\xi_0$ that splits the locations of the two overlapping
1-vortices, whose form is
\begin{equation}
\xi_0(\vec{x})= \left(r h_{20}(r) \sin \theta \, , \hspace{0.4cm}
r h_{20}(r) \cos \theta \, , \hspace{0.4cm} -\frac{r
h'_{20}(r)}{f_2(r)} \, , \hspace{0.4cm} 0 \right)^t \, ,
\label{zeromode}
\end{equation}
where $h_{20}(r)$ can be obtained numerically\footnote{Note, (\ref{zeromode})
separates the 1-vortices along the $y$-axis. Obviously, this has
no essential consequence for the calculations shown here because of the
rotational symmetry.} (see Section 4.1 in ref.\cite{AlGarGuil2}),
and note that $h_{20}(r)$ is a decreasing function so
$h_{20}'(r)<0$. Therefore, the expansion (\ref{expansion1}) simplifies to
\begin{equation}
\widetilde{\psi}^{(2)}(\vec{x}) = f_2(r) e^{2i\theta} - \epsilon \,
\frac{r h_{20}'(r)}{f_2(r)}\,  + \cdots \, .
\label{expansion1b}
\end{equation}
A relation between the perturbation parameter $\epsilon$ and the
small distance parameter $d$ can be derived from the zeros of
(\ref{expansion1b}). Near the origin, $f_2(r) \approx d_2 r^2$ and
$h_{20}(r) \approx 1 + c_2^{(2,0)}r^2$ with $d_2\approx 0.236146$
and $c_2^{(2,0)}\approx -0.277308$. Therefore
\begin{equation}
\epsilon = \frac{(d_2)^2}{2 \, |c_2^{(2,0)}|} \, d^2 \, .
\label{epsilondrel}
\end{equation}
Notice that (\ref{expansion1b}) is a first-order expansion in
$\epsilon$, so we are restricted here to a first-order treatment of
the spectral problem (\ref{pdeA}). A higher-order expansion is a
non-trivial task beyond the scope of this paper.

\begin{figure}[h]
	\centering
	\includegraphics[height=3.5cm]{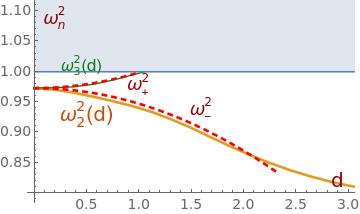}  \hspace{1cm}
	\includegraphics[height=3.5cm]{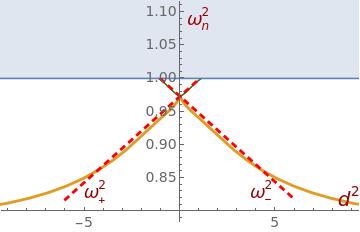}
	\caption{Comparison between the numerical results (solid) and
          first-order perturbative results for small $d$ (dashed), for the
          eigenvalues $\omega_2^2$ and $\omega_3^2$: (left) plotted
          against $d$, (right) plotted against $c=d^2$.}
        \label{fig:spectrum2analytic}
\end{figure}

In addition to (\ref{expansion1}), the eigenmodes
and eigenvalues must be expanded as
\begin{eqnarray}
&& \widetilde{a}_1(\vec{x}) = a_{1}(\vec{x}) + \epsilon \,
\delta a_{1}(\vec{x}) + \cdots \, , \nonumber \\ 
&& \widetilde{\omega}^2 = \omega^2 + \epsilon \, \delta\omega^2 +
\cdots \, , 
\end{eqnarray}
where $a_{1}(\vec{x})$ and $\omega^2$ are a joint eigenmode and
eigenvalue of the fluctuation operator ${\cal H}^-$ associated to the
rotationally-invariant 2-vortex, and $\epsilon$ is related to $d$ as
in (\ref{epsilondrel}). At first order, eq.(\ref{pdeA}) reduces to
\begin{equation}
\left[- \nabla^2  + |\psi^{(2)} (\vec{x}) |^2 -\omega^2 \right]
\delta a_1(\vec{x}) = \left( \delta\omega^2 - 2 \, {\rm Re} [
  \overline{\psi}^{(2)} (\vec{x}) \, {\delta\psi}^{(2)} (\vec{x}) ]
\right) a_1(\vec{x}) \, .
\label{firstorder}
\end{equation}
From the Fredholm alternative it is known that the projection
of the right-hand side of (\ref{firstorder}) on to the
homogeneous solution (the eigenmode $a_1(\vec{x})$ at $d=0$) must
be zero in order to obtain bounded solutions. This implies that
\begin{equation}
\delta\omega^2 = \frac{\int d^2 x \, 2 \, {\rm Re} [
  \overline{\psi}^{(2)} (\vec{x}) \, {\delta\psi}^{(2)} (\vec{x}) ] \,
  (a_1(\vec{x}))^2}{\int d^2 x \, (a_1(\vec{x}))^2} \, .
\label{omega1}
\end{equation}
We can now evaluate the integrals in (\ref{omega1}) to estimate the
three eigenvalues arising for small $d$.
\begin{enumerate}
\item For the perturbed $k=0$ mode at $d=0$,
\begin{equation}
\delta\omega^2 = \frac{\int_0^\infty dr  \, 2r^2 |h_{20}'(r)|  \,
  v_{20}^2(r) \int_0^{2\pi} d\theta \, \cos 2\theta }{2\pi
  \int_0^\infty dr  \, r v_{20}^2(r)}=0
\end{equation}
because the angular integral vanishes. Therefore, the lowest
non-degenerate eigenvalue has no quadratic dependence on $d$,
so the leading dependence is quartic (at least),
$\widetilde{\omega}_1^2 = \omega_1^2 + O(d^4)$.

\item For the $k=1$ mode proportional to $\cos \theta$ at
$d=0$,
\begin{equation}
\delta\omega^2 = \frac{\int_0^\infty dr  \, 2r^2 |h_{20}'(r)|  \,
  v_{21}^2(r) \int_0^{2\pi} d\theta \, \cos 2\theta \cos^2 \theta }{
  \int_0^\infty dr  \, r v_{21}^2(r) \int_0^{2\pi} d\theta \cos^2
  \theta} = \frac{\int_0^\infty dr   \,r^2 |h_{20}'(r)|  \,
  v_{21}^2(r)}{ \int_0^\infty dr  \, r v_{21}^2(r)} \approx
  0.257323 \, .
\end{equation}

\item For the $k=1$ mode proportional to $\sin \theta$ at $d=0$,
\begin{equation}
\delta\omega^2 = \frac{\int_0^\infty dr  \, 2r^2 |h_{20}'(r)|  \,
  v_{21}^2(r) \int_0^{2\pi} d\theta \, \cos 2\theta \sin^2 \theta }{
  \int_0^\infty dr  \, r v_{21}^2(r) \int_0^{2\pi} d\theta \sin^2
  \theta} = -\frac{\int_0^\infty dr   \, r^2 |h_{20}'(r)|  \,
  v_{21}^2(r)}{ \int_0^\infty dr  \, r v_{21}^2(r)} \approx
  -0.257323 \, .
\end{equation}

\end{enumerate}
Combining the last two results, we find the splitting of the
eigenvalues $\omega_3^2$ and $\omega_2^2$ that are degenerate at $d=0$,
\begin{equation}
  \widetilde{\omega}_{3,2}^2 \equiv \widetilde{\omega}_{\pm}^2
  \approx 0.97303 \pm \frac{(0.236146)^2}{2 \cdot
  0.277308} \, 0.257323 \, d^2  \approx 0.97303 \pm 0.025873\, d^2 \, ,
\label{omegapm}
\end{equation}
as shown in Figure \ref{fig:spectrum2analytic} (left). The similar
dependence on $d^2$, apart from the sign, is striking, and will be
clarified in the next Section.

\section{Insight from the 2-vortex moduli space} 

The work of Taubes on BPS $n$-vortex solutions \cite{Tau} and the
later work of Samols on the Riemannian geometry of the $n$-vortex
moduli space \cite{Samols}, reviewed in \cite{ManSut}, make
clear that it is best to set $z= x+ iy$, and to use
the complex variable $Z = X+iY$ to denote the location of a vortex (a
zero of the scalar field), instead of the Cartesian 2-vector $(X,Y)$.
An $n$-vortex solution is characterised by its $n$ unordered zeros;
good complex coordinates on the moduli space are therefore the $n$
elementary symmetric polynomials in these zeros. If the zeros are at
$\{Z_1,Z_2,\dots,Z_n\}$, these coordinates are (up to sign) the
coefficients of the polynomial
\begin{equation}
P(z) = (z-Z_1)(z-Z_2)\dots(z-Z_n) \, .
\end{equation}
Taubes showed that an $n$-vortex with these zeros exists and is unique up to
gauge transformations, and in a convenient gauge has a scalar field $\phi$
that is a product of a real function with the polynomial $P(z)$.

For a 2-vortex with zeros at $Z_1$ and $Z_2$,
\begin{equation}
P(z) = z^2 - (Z_1+Z_2)z + Z_1Z_2 \, ,
\end{equation}
so good, 2-vortex moduli space coordinates are the centre of mass
$\frac{1}{2}(Z_1+Z_2)$ and the product $-Z_1Z_2$
(this sign choice is convenient). We are interested in
vortices with centre of mass at the origin. In particular, if the
vortices have Cartesian locations $(d,0)$ and $(-d,0)$, as above, then
$Z_1 = d$ and $Z_2 = -d$, and the good coordinate for these centred
vortices is $c = -Z_1Z_2 = d^2$. When $c=0$, the vortices coincide at
the origin, and the 2-vortex is rotationally invariant. Samols showed
that the metric on the moduli space of centred 2-vortices has the
form $ds^2 = F(|c|) \, dc \, d{\bar c}$, with $F$ smooth and positive,
including at $c=0$.

There is a geodesic motion through moduli space, where $c$ moves
smoothly along the real axis from positive to negative
values and the velocity of $c$ remains negative throughout. What this
means is that the vortices scatter through a
right angle. If the vortices have Cartesian locations $(0,d)$ and
$(0,-d)$, then $Z_1 = id$ and $Z_2 = -id$, so $c = -Z_1Z_2 = -d^2$. When
$c$ is positive, the vortex locations are on the $x$-axis; when
$c$ is negative, they are on the $y$-axis.

The eigenvalues and eigenmodes of the fluctuation operator around a
2-vortex are expected to flow smoothly over the moduli space. In
particular, for centred vortices moving on the $x$-axis, the flow is
smooth as a function of $c = d^2$. $c$ is a better
coordinate than $d$, and the flow remains smooth if the range
of $c$ is extended to negative values, corresponding to a right-angle
scattering of the vortices. Furthermore, there is a symmetry between
the vortex configuration with modulus $c$, and with modulus $-c$; they
differ by a rotation through a right angle. So the eigenvalues of the
discrete modes are the same at $c$ and $-c$, and the eigenmodes
should be related by a right-angle rotation.

\begin{figure}[h]
	\centering
	\includegraphics[height=4.5cm]{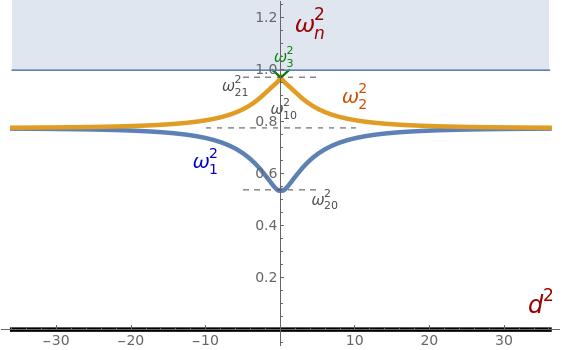} \hspace{0.5cm}
	\includegraphics[height=4.5cm]{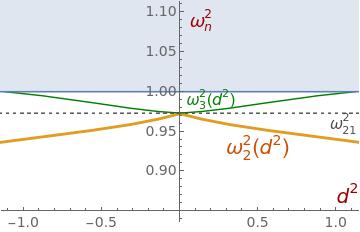}
	\caption{Dependence of the 2-vortex fluctuation
          operator eigenvalues on the parameter $c=d^2$ (left). The
          eigenvalue crossover region in the left figure has been
          enlarged on the right.}
        \label{fig:spectrum2v03}
\end{figure}

The results we have obtained show that these expectations are fulfilled.
The eigenvalues as a function of $d$ (for $d \ge 0$) are shown in Figure
\ref{fig:spectrum2}, but additionally, we have recalibrated the axes to
show their dependence on $c=d^2$, and have reflected the graphs
from right-to-left to show the eigenvalues for negative $c$. The
outcome is Figure \ref{fig:spectrum2v03}. The same procedure has been
applied to the perturbative results, as shown in Figure
\ref{fig:spectrum2analytic} (right). The features to note in
Figures \ref{fig:spectrum2}, \ref{fig:spectrum2analytic}
and \ref{fig:spectrum2v03} are (1) the graphs are smooth;
(2) for small $c$, the dependence of the lowest eigenvalue $\omega_1^2$ on
$c$ appears to be quadratic, as expected for a smooth symmetric function
with a minimum. This quadratic dependence was verified through the
perturbative analysis of Section \ref{Pert}. It means that the
dependence of $\omega_1^2$ on
$d$, shown in Figure \ref{fig:spectrum2}, is quartic, something that
would be rather curious if one did not take into account the moduli space
geometry; (3) the two eigenvalues $\omega_2^2$ and $\omega_3^2$ cross over
smoothly and linearly at $c=0$. This happens because for $c > 0$, the
eigenmode for $\omega_2^2$ is antisymmetric in $x$ and symmetric in $y$,
whereas for $\omega_3^2$ it is the opposite way. As the vortices scatter
through a right angle, the symmetry axes are exchanged, and the
eigenvalues exchange their order. The linear dependence of these
eigenvalues on $c=d^2$, and the smooth crossover, is verified by the
formula (\ref{omegapm}) obtained through the perturbative analysis. In terms of
$d$, the eigenvalues have quadratic dependence near $d=0$, and
crucially, the coefficients of the quadratic terms have opposite signs.
The numerical results, shown in Figure \ref{fig:spectrum2analytic}, do
not exactly match these expectations. This could be because the
numerical analysis is tricky for $d$ very small, or it could be
because the locations of the vortex zeros have moved slightly during
the relaxation of the vortex configuration from the initial ansatz
(\ref{productansatz}) to the optimised solution. 

Recall that eigenvalue crossing is not generic as one parameter varies;
instead the eigenvalues tend to repel and avoid crossing. However,
here the crossing eigenvalues have eigenmodes with opposite
symmetries, so there is no repulsion.

\section{Outlook}

We have obtained some detailed understanding of the discrete
shape modes of BPS 2-vortex solutions, and how they vary with
the separation of the 1-vortex constituents. It would be interesting
to study the modification to the low-energy scattering of vortices when
these modes are excited, either classically or quantum mechanically.
The geodesic motion through the 2-vortex moduli space \cite{Samols} will
be supplemented by oscillation of the discrete modes and some
potential energy function on the moduli
space. An adiabatic treatment should be possible if the energy in the
modes of oscillation is comparable with the kinetic energy of the
translational motion of the vortices, and both are small. The simplest
case would be when just one shape mode is excited. As the
shape modes have different symmetries, the transfer of energy from one
mode to another is likely to be suppressed. A complication may occur
at the critical separation where one mode enters the continuum, should
that mode be excited. We have seen that if vortices approach from a
large distance, and their radial shape modes are excited in counterphase,
then after they approach and scatter through a right angle, then it is
this mode that enters the continuum.   

A related program would be a search for fermionic bound states to
vortices and an examination of their properties. This involves
replacing Weinberg's first-order differential operator by the Dirac
operator, and investigating the spectral problem

\begin{equation}
\left\{ -i \gamma^j \left( \frac{\partial}{\partial x_j}-i
    V_j(x_1,x_2)\right)+2 \vert
  \psi(x_1,x_2)\vert\right\}\Psi(x_1,x_2;\omega)=\omega
\Psi(x_1,x_2;\omega) \, ,
\end{equation}
where $\gamma^0, \gamma^1, \gamma^2$ are $2 \times 2$ matrices that
generate the Clifford Algebra of $\mathbb{R}^{1,2}$:
$\{\gamma^\mu,\gamma^\nu\}=2 g^{\mu\nu}$. The 2-component spinors
$\Psi$ belong to the fundamental representation of the group
$\bf{Spin}(2,1;\mathbb{R})$, the component of the double cover of
the Lorentz group $\mathbb{S}\mathbb{O}(2,1;\mathbb{R})$ connected
to the identity. Its group elements and Lie algebra are given by 
\begin{equation}
S(\omega_{\mu\nu})={\rm exp}\left[\frac{i}{2} \omega_{\mu\nu}
\sigma^{\mu\nu} \right] \, , \quad \omega_{\mu\nu}=-\omega_{\nu\mu}
\, , \quad \sigma^{\mu\nu}=\frac{1}{2}[\gamma^\mu,\gamma^\nu] \, .
\end{equation}
Finally, we mention that a similar analysis may be performed for the
BPS vortices in the gauged massive non-linear sigma model, discussed
in refs.\cite{Schroers, Nitta, Fuertes, Speight2021} for example.
Although such models are non-renormalizable, they may arise as
low-energy effective theories within non-Abelian gauge theory or
even string theory.

\section*{Acknowledgments}

This work developed from a presentation at the SIG XI workshop, Jagiellonian
University, Krakow. We thank A. Wereszczinski, K. Oles and C. Naya
Rodriguez for organising the workshop.

This research was supported by the Spanish MCIN with funding from European
Union NextGenerationEU (PRTRC17.I1) and Consejeria de Educacion from
JCyL through the QCAYLE project, as well as the MCIN project
PID2020-113406GB-I0. This research has made use of the
high-performance computing resources of the Castilla y Le\'on
Supercomputing Center (SCAYLE), financed by the European Regional
Development Fund (ERDF). NSM is partially supported by consolidated
grant ST/T000694/1 from the UK STFC.

\end{document}